# Pressure-induced superconductivity in layered pnictogen diselenide $NdO_{0.8}F_{0.2}Sb_{1-x}Bi_xSe_2$ ($x$ = 0.3 and 0.7)

Ryo Matsumoto,[1,2] Yosuke Goto,[3,*] Sayaka Yamamoto,[1,2,4] Kenta Sudo,[3,#] Hidetomo Usui,[5] Akira Miura,[6] Chikako Moriyoshi,[7] Yoshihiro Kuroiwa,[7] Shintaro Adachi,[1] Tetsuo Irifune,[8] Hiroyuki Takeya,[1] Hiromi Tanaka,[4] Kazuhiko Kuroki,[9] Yoshikazu Mizuguchi,[3] and Yoshihiko Takano[1,2]

[1]*National Institute for Materials Science, 1-2-1 Sengen, Tsukuba, Ibaraki 305-0047, Japan*

[2]*University of Tsukuba, 1-1-1 Tennodai, Tsukuba, Ibaraki 305-8577, Japan*

[3]*Department of Physics, Tokyo Metropolitan University, 1-1 Minami-osawa, Hachioji, Tokyo 192-0397, Japan*

[4]*National Institute of Technology, Yonago College, 4448 Hikona, Yonago, Tottori 683-8502, Japan*

[5]*Department of Physics and Materials Science, Shimane University, Matsue, 690-8504, Japan*

[6] *Faculty of Engineering, Hokkaido University, Kita 13, Nishi 8 Sapporo 060-8628, Japan*

[7] *Department of Physical Science, Hiroshima University, 1-3-1 Kagamiyama, Higashihiroshima, Hiroshima 739-8526, Japan*

[8]*Geodynamics Research Center, Ehime University, Matsuyama, Ehime 790-8577, Japan*

[9]*Department of Physics, Osaka University, Toyonaka, Osaka 560-0043, Japan*

[#]*Present address: High Field Laboratory for Superconducting Materials, Institute for Materials Research, Tohoku University, Sendai 980-8577, Japan*

*E-mail: y_goto@tmu.ac.jp

**Abstract**

Polycrystalline samples of layered pnictogen (Pn) diselenide $NdO_{0.8}F_{0.2}Sb_{1-x}Bi_xSe_2$ ($x$ = 0 to 0.8) were successfully synthesized by solid-state reactions. Electrical resistivity in the synthesized samples systematically decreases with an increase in Bi content $x$, accompanied by an insulator to metal-like transition at $x$ = 0.5–0.6. Crystal structure analysis using synchrotron X-ray diffraction suggests that this transition correlates with anomalous change in $c$-axis length and/or distortion in Pn–Se conducting plane. Emergence of pressured-induced superconductivity was investigated using diamond anvil cell (DAC) with boron-doped diamond electrodes, employing $x$ = 0.3 and 0.7 as the representative samples. For Sb-rich one ($x$ = 0.3), we observed a signature of superconducting transition with $T_c^{onset}$ = 5.15 ± 0.05 K at 50 GPa, which is the first-ever report of the superconductivity in layered $SbCh_2$-based (Ch: chalcogen) compounds. The $T_c^{onset}$ of $x$ = 0.3 increased with increasing pressure and reached 7.65 ± 0.05 K at 70.8 GPa, followed by the gradual decrease in $T_c$ up to 90 GPa. For Bi-rich one ($x$ = 0.7), a superconducting transition with $T_c^{onset}$ = 5.5 ± 0.6 K was induced at 43.5 GPa, which is the almost comparable to that of $x$ = 0.3; besides, upper critical field ($H_{c2}$) is evaluated to be ~10 T for $x$ = 0.7, which is higher than that of $x$ = 0.3 ($H_{c2}$ = 6.7 T at 50 GPa).

## 1. Introduction

Since the discovery of the $BiCh_2$-based (Ch: S, Se) layered superconductors in 2012, such as $Bi_4O_4S_3$ and R(O,F)$BiS_2$ (R: rare-earth element), this family of compounds has received much attention as a new class of layered superconductors [1-8]. The crystal structure is composed of alternate stacks of electrically conducting $BiCh_2$ layers and insulating (blocking) carrier reservoir layers, as schematically depicted in Fig. 1. Several types of $BiCh_2$-based superconductors have been reported, and the highest-record $T_c$ of 11 K was obtained for $LaO_{0.5}F_{0.5}BiS_2$ using high-pressure technique [3,9]. In the early stages of research on a paring mechanism of $BiCh_2$-based compounds, it was suggested that the conventional superconductivity with a fully gapped *s*-wave state is realized, on the basis of the first-principles calculation [10], Raman scattering [11], muon-spin spectroscopy [12], and thermal conductivity [13] experiments. However, recent first-principles calculations [14], angle-resolved photoemission spectroscopy [15], and Se isotope effect [16] proposed the possibility of unconventional pairing mechanisms in $BiCh_2$-based superconductors.

Generally speaking, the parent phase of $BiCh_2$-based compounds is band gap insulators. These compounds show superconductivity when electron carriers are generated at the conduction bands. Because the conduction bands are mainly composed of hybridization between Bi 6*p* and Ch *p* orbitals [17,18], it is quite reasonable to expect that the understanding of the isovalent doping effects of Sb for Bi is essentially important to elucidate the superconducting mechanisms and the underlying physics in this family of compounds. In addition, Sb doped $BiCh_2$-based compounds, namely $SbCh_2$-based compounds, are also attractive from the viewpoint of the thermoelectric and topological materials [19-21]. However, most of $SbCh_2$-based compounds have been insulators so far, in spite of an attempts to dope electron carriers [22-26]. Although the origin of insulating nature in $SbCh_2$-based compounds has not been fully understood yet, it is probably due to insufficient orbital overlapping between Sb 5*p* and Ch *p* orbitals [27]. This scenario is supported by the observation of insulator to metal transition for $Ce(O,F)SbS_2$ under high pressure, which promotes orbital overlapping [25].

Recently, we have reported the effects of Bi doping on high-temperature thermoelectric transport properties of $SbSe_2$-based layered compounds $NdO_{0.8}F_{0.2}Sb_{1-x}Bi_xSe_2$ ($x \leq 0.4$) [26]. Room temperature electrical resistivity was decreased down from 300 mΩcm for $x = 0$ to 8 mΩcm for $x = 0.4$. However, the temperature dependence of electrical resistivity was still insulating for $x \leq 0.4$, in spite of F doping content is 20%. In the present study, we show that Bi-rich composition up to ca. $x = 0.8$ can be obtained by optimizing synthesis temperature. Electrical resistivity in the synthesized samples was systematically decreased with an increase in Bi content $x$, accompanied by an insulator to metal-like transition at $x = 0.5$–$0.6$. Electrical transport of $x = 0.3$ and 0.7 was examined under high pressure exceeding 50 GPa using diamond anvil cell (DAC) with boron-doped diamond electrodes. It was found that both samples show resistance drop at low temperature, which is a signature of superconductivity in these compounds.

## 2. Method

### 2.1 Sample preparation

Polycrystalline samples of $NdO_{0.8}F_{0.2}Sb_{1-x}Bi_xSe_2$ ($x$ = 0, 0.1, 0.2, 0.3, 0.4, 0.5, 0.6, 0.7, 0.8) were prepared by solid-state reactions using dehydrated $Nd_2O_3$, NdSe, $NdSe_2$, Sb (99.9%), Bi (99.999%), and Se (99.999%) as starting materials. The dehydrated $Nd_2O_3$ was prepared by heating commercial $Nd_2O_3$ powder (99.9%) at 600 °C for 10 h in air. To obtain the NdSe and $NdSe_2$ mixtures, Nd (99.9%) and Se in a molar ratio of 2:3 were heated at 500 °C for 10 h in an evacuated silica tube. Because the Nd powder is reactive in air and a moist atmosphere, this process was carried out in an Ar-filled glovebox with a gas-purifier system. Then, a stoichiometric mixture of these starting materials was pressed into a pellet and heated for 15 h at 700 °C for $x \leq 0.4$ and at 650 °C for $x \geq 0.5$ in an evacuated silica tube. The obtained sample was ground, mixed, pelletized, and heated with the same heating condition. It should be noted that lowered synthesis temperature (650 °C) is employed for $x \geq 0.5$, because amounts of impurity phases, $Bi_2Se_3$ and $Nd_2O_2Se$, was significant when these samples were synthesized at 700 °C.

## 2.2 Characterization

The chemical compositions of the obtained samples were examined using an energy dispersive X-ray spectrometer (EDX; Oxford, SwiftED3000). The phase purity and crystal structure of the samples were examined by synchrotron powder X-ray diffraction (SPXRD) performed at the BL02B2 beamline of SPring-8 (proposal numbers 2018A0074 and 2018B1246). The measurements were performed at 297 K. The diffraction data were collected using a high-resolution one-dimensional semiconductor detector (multiple MYTHEN system) [28]. The wavelength of the

radiation beam was determined to be 0.495274(1) Å (proposal number 2018A0074) and 0.496345(1) Å (No. 2018B1246) using a $CeO_2$ standard. The crystal structure parameters were refined by the Rietveld method using RIETAN-FP software [29]. The crystal structure was visualized using VESTA software [30].

### 2.3 Transport measurement

Electrical resistivity at ambient pressure was measured using the four-probe method on a GM refrigerator system. Measurements of electrical resistance under high pressure were performed on polycrystalline powder using an originally designed DAC with boron-doped diamond electrodes [25,31,32] on the bottom anvil of nanopolycrystalline diamond [33], as shown in Fig. 2. The sample was placed on the boron-doped diamond electrodes in the center of the bottom anvil. The surface of the bottom anvil except for the sample space and electrical terminal were covered by the undoped diamond insulating layer. The cubic boron nitride powders with ruby manometer were used as a pressure-transmitting medium. The applied pressure ($P$) was estimated by the fluorescence from ruby powders [34] and the Raman spectrum from the culet of top diamond anvil [35] by an inVia Raman Microscope (RENISHAW). The resistance was measured by a standard four-probe method on a Physical Property Measurement System (Quantum Design: PPMS).

### 2.4 First-principles calculation

Density functional theory calculations were performed using VASP software [36-39]. This code implements the projector-augmented wave method [40]. The electronic structures were self-consistently calculated within the generalized gradient approximation proposed by Perdew, Burke, and Ernzerhof [41]. Here, we set a 15×15×5 k-mesh and cut-off energy of 550 eV with the inclusion of spin-orbit coupling. The virtual crystal approximation was used to take into account the effect of partial substitution of F for O and Bi for Sb, respectively. For the calculations, we adopted the experimentally determined crystal structures described in Table S1 in the Supplemental Materials [42].

## 3. Results and discussion

### 3.1 Crystal structure and chemical composition

Figure 3 shows the SPXRD pattern and Rietveld fitting results for $x$ = 0.7 as a representative data. Almost all the diffraction peaks can be assigned to those of the tetragonal *P4/nmm* space group, indicating that the obtained sample is mainly composed of $NdO_{0.8}F_{0.2}Sb_{1-x}Bi_xSe_2$-type phase. However, several peaks attributable to impurity phases, $Nd_2O_2Se$ (4.1 wt%), NdOF (1.9 wt%), and $Bi_2Se_3$ (0.8 wt%), were also observed. Most of the diffraction peaks for other samples can also be assigned to those of the tetragonal *P4/nmm* space group, as shown in Fig. S1. Amounts of impurity phases are significant for $x$ = 0.8, $Nd_2O_2Se$ (13.2 wt%) and $Bi_2Se_3$ (12.9 wt%), as summarized in Fig. S2, suggesting solubility limit of Bi for Sb in $NdO_{0.8}F_{0.2}Sb_{1-x}Bi_xSe_2$ is at around this composition. Figure 4(a) shows the chemical composition ratios of Nd, Sb, Bi, and Se determined using EDX. The results indicate that the chemical compositions of the obtained samples are in reasonable agreement with the nominal compositions of the starting materials.

Figure 4(b) depicts the calculated lattice parameters. The lattice parameter *a* increased almost linearly with increasing *x* owing to the larger ionic radius of Bi ions than that of Sb ions (Shannon's five-coordinate ionic radius, $r_{Bi}^{3+}$ = 96 pm and $r_{Sb}^{3+}$ = 80 pm) [43]. On the other hand, the *c* exhibits anomalous change at around *x* = 0.5–0.6. Namely, *c*-axis length exhibits longer value than that expected from linear trend. Although the origin of this deviation from linear relationship between *c* and *x* is not clear yet, it seems to be correlated to electrical carrier transport of the present samples, as described below. Note that *c*-axis length tends to correlate with amount of electron doping in the $BiCh_2$-based systems [2-8].

Figures 5(a)-(d) show selected bond distances and angle. Bond distances for in-plane Bi/Sb (Pn)-Se1 and Pn-Se2 increased almost linearly with increasing *x*, consistent with an increase in lattice parameters. On the other hand, interplane Pn-Se1 distance is in 3.43 ± 0.03 Å for all of examined samples. As a result, Se1-Pn-Se1 bond angle tends to decrease with increasing *x*. This indicates that Pn–Se1 plane is distorted by Bi doping. The Se1-Pn-Se1 bond angle can be classified into two groups, namely, Se1-Pn-Se1 bond angle of ca. 171° for Sb-rich composition ($x \leq 0.3$), and of ca. 169° for Bi-rich composition ($x \geq 0.6$). As shown below, the temperature dependence of electrical resistivity near room temperature turns from insulating to metallic behavior at the intermediate region, *x* = 0.5–0.6. It should be noted that the distortion of Pn–Se1 plane by applying chemical pressure is similar to the case of $LaO_{0.5}F_{0.5}Bi(S,Se)_2$ [27].

### 3.2 Transport properties under ambient pressure

Figure 6(a) shows the temperature dependence of electrical resistivity measured under ambient pressure. For $x = 0$, electrical resistivity increased with decreasing temperature, leading to $\sim 10^5$ Ωcm at 2 K. As shown in Fig. 6(b), room temperature electrical resistivity tends to decrease with increasing $x$, in spite of isovalent doping of Bi for Sb. This is most likely due to increased overlapping between Sb/Bi p and Se p orbitals, so called in-plane chemical pressure effects, as established in $BiCh_2$-based compounds [27]. As shown in Fig. 6(c), temperature dependence of electrical resistivity turns from insulating to metallic-like behavior at $x = 0.5$–$0.6$, in which anomalous change of $c$-axis length and Se1-Pn-Se1 bond angle is observed (see Fig 4(b) and 5(d)). Notably, weakly localized (insulating) behavior is still observed at low temperature for all samples. For example, electrical resistivity increases with decreasing temperature below 20 K for $x = 0.8$. We note that transport measurements were performed on polycrystalline samples in this study. Usually, the transport properties of the polycrystalline sample have some uncertainty due to the presence of, e.g. the impurity phase, grain boundary, and anisotropy. Our samples have a purity of approximately >94% on the basis of Rietveld analysis except for $x = 0.4$ and 0.8, as summarized in the Fig. S2. Typical impurity phases are NdOF, $Sb_2Se_3$, and $Bi_2Se_3$, which are known as semiconductor or insulator at ambient conditions. Therefore, electrical conductivity is dominated by $NdO_{0.8}F_{0.2}Sb_{1-x}Bi_xSe_2$. Regarding grain boundary, it is reasonable to expect that the effects on electrical resistivity is temperature independent. At the same time, it seems to be reasonable that anisotropy is almost unchanged by doping, because the present samples are isostructural compounds. Therefore, it is safely concluded that transition from insulator to metal/superconductor is intrinsic properties of $NdO_{0.8}F_{0.2}Sb_{1-x}Bi_xSe_2$, in spite of polycrystalline nature of the samples.

**3.3 Transport properties under high pressure**

To investigate the electrical transport under high pressure, we employed $x$ = 0.3 and 0.7 as the representatives of Sb-rich and Bi-rich composition, respectively. It should be noted that $x$ = 0.8 is ruled out for high pressure measurements because of the existence of nonnegligible amount of $Bi_2Se_3$ secondary phase (12.9 wt%), which shows superconductivity under high pressure [44,45]. $Bi_2Se_3$ content in $x$ = 0.3 and 0.7 was evaluated to be 0.4 and 0.8 wt%, respectively.

Figure 7 shows the temperature dependences of electrical resistance for $x$ = 0.3 under various pressures (a) from 12.2 GPa to 50 GPa, and (b) from 50 GPa to 90 GPa. It is difficult to measure $P$ = 0 data because the contact between the sample and electrodes are kept by a pressure welding. The resistance decreases with increasing pressure. Although the data under 23.7, 29.1, and 36.4 GPa were noisy probably due to a high contact resistance at low temperature region, the contact was improved by an increase of the pressure. A sudden drop of resistance was observed at 5.15 ± 0.05 K at 50 GPa, corresponding to an indication of a superconducting transition. The criterion for $T_c$ is described below. At 57.2 GPa, the resistance at 10 K decreases about four orders in magnitude as compared to that at 12.2 GPa, showing the insulator to metal transition at this pressure. Broadening features of the resistance curve below 50 K at 57.2 GPa and 64.1 GPa were observed. Although the reason of this broad feature is still unclear, at least, it could not be attributed to superconductivity because the normal resistance above 7-8 K was independent of the applied magnetic field, as later shown in Fig. 8. The resistance continued to decrease up to 90 GPa, and then, the diamond anvil was broken. It should be noted that zero resistivity was not observed at low temperature, most likely due to inhomogeneity of applied pressure for polycrystalline sample, which is generally occurred with DAC systems [46].

Figure 7(c) shows resistance below 10 K for $x$ = 0.3 under various pressures from 50 GPa to 90 GPa. In the present study, we evaluated $T_c$ as follows: (1) $T_c^{upper}$ was determined from a branch point on the differential curve under 0 T and 7 T (or sufficiently high magnetic fields), as shown in Fig. S3(a). (2) $T_c^{lower}$ was determined from an intersection point between the strait line of the normal resistance region and extended line from resistance after superconducting transition under 0 T and 7 T (or sufficiently high magnetic fields, see Fig. S3(b)). (3) $T_c^{onset}$ was determined from the average value of $T_c^{upper}$ and $T_c^{lower}$. (4) These three kinds of $T_c$ value ware used for the error bars. We first observed the superconducting transition with $T_c^{onset}$ = 5.15 ± 0.05 K under 50 GPa. The $T_c^{onset}$ was enhanced with an increase of applied pressure up to 70.8 GPa, reaching the maximum $T_c^{onset}$ of 7.65 ± 0.05 K. Under further compression, the $T_c^{onset}$ was gradually decreased.

To confirm that the observed drop of resistance originates from the superconductivity, we measured the resistance under magnetic fields. Figure 8 shows temperature dependences of resistance for $x$ = 0.3 in magnetic fields under representative pressures of (a) 50 GPa, (b) 70.8 GPa, and (c) 90 GPa. The drops of resistance are gradually suppressed by an increase of applied magnetic field, indicating the drops of resistance come from the superconductivity. The insets show the temperature dependences of upper critical field $H_{c2}$ estimated from the Werthamer-Helfand-Hohenberg (WHH) approximation for type II superconductors in a dirty limit [47]. The $H_{c2}(0)$ was evaluated to be 6.7 T, 5.6 T, and 4.6 T at 50 GPa, 70.8 GPa, and 90 GPa, respectively, which are comparable to that of $BiCh_2$-based superconductors.

Figure 9(a) shows temperature dependences of resistance in Bi-rich sample, $x = 0.7$, under pressures from 0.6 GPa to 52 GPa. An insulating behavior was observed up to 36.5 GPa, while resistivity of this composition measured at ambient pressure shows metallic behavior at temperatures between 100 and 300 K (see Fig. 6(c)). This is probably due to weak coupling at grain boundary, because the measurements under high pressure are performed on polycrystalline powder samples. We can see the hump-like broad anomaly around 200 K under 0.6 GPa to 36.5 GPa. This anomaly is suppressed by increasing pressure up to 43.5 GPa, where the resistance drop corresponding to the superconducting transition was observed at 5.5 ± 0.6 K. This is quite similar to the case of pressure-induced superconductivity in $EuFBiS_2$ [48], although the origin of the anomaly around 200 K is still unclear. For $x = 0.7$, the critical parameters to induce superconductivity ($T_c$ and $P_c$) are almost comparable to those of $x = 0.3$. Figures 9(b) and 9(c) show the temperature dependences of resistance in $x = 0.7$ under magnetic fields at (b) 43.5 GPa and (c) 52.0 GPa. Again, suppression of drops in resistance by applying magnetic field indicate that this corresponds to a superconducting transition. The $H_{c2}(0)$ of $x = 0.7$ was evaluated to be 10.9 T and 9.4 T at 43.5 GPa and 52.0 GPa, respectively. These are distinctly higher than that of $x = 0.3$.

Figures 10(a) and 10(b) show pressure-phase diagrams for $T_c^{onset}$ and resistances at 200 K and 10 K for $x$ = 0.3 and 0.7. In spite of different Sb/Bi substitutional ratio, both compositions seem to exhibit quite similar responses against the applied pressures. Namely, critical pressure to induce superconductivity is as high as ~45 GPa for both the Sb rich ($x$ = 0.3) and Bi rich ($x$ = 0.7) composition. This seems to contradict our simple expectation that superconductivity is more easily induced under high pressure in Bi rich sample, because of its metallic-like nature at ambient pressure. On the other hand, $H_{c2}$ is distinctly different in these samples, $H_{c2}$ = 6.7 T for $x$ = 0.3 at 50 GPa and $H_{c2}$ = 10.9 T for $x$ = 0.7 at 43.5 GPa.

Here, let us emphasize again that zero resistivity was not obtained in the present study, probably because of inhomogeneity of applied pressure for polycrystalline sample. Single crystals will be helpful to observe a sharp superconducting transition, as well as to investigate the anisotropy in physical properties, such as normal state $\rho$, $P_c$, $H_{c2}$, and so forth. However, a possibility of resistivity drop from $Bi_2Se_3$ secondary phase, which shows superconductivity under high pressure [44,45], can be ruled out on the basis of following reasons: (1) We employed $x$ = 0.3 and 0.7 for high pressure measurements. In these samples, the amount of $Bi_2Se_3$ is evaluated to be 0.4 wt% for $x$ = 0.3 and 0.8 wt% for $x$ = 0.7. (2) Critical pressure ($P_c$) to induce superconductivity for the present samples are ~ 50 GPa, which is distinctly higher than that reported for $Bi_2Se_3$ (14–25 GPa at 2 K). (3) Upper critical field ($H_{c2}$) of our sample reaches 6.7 T for $x$ = 0.3 and 10.9 T for $x$ = 0.7. These are higher than $H_{c2}$ reported on $Bi_2Se_3$ under high pressure, 4.5 T. If superconducting path in our samples is created due to $Bi_2Se_3$, $H_{c2}$ should be less than 4.5 T because of its filamentary nature.

### 3.4 First-principles calculations

To discuss the electronic structure of $NdO_{0.8}F_{0.2}Sb_{1-x}Bi_xSe_2$, we performed the first-principles calculations using VASP software. Figure 11 shows band dispersion of $x = 0$, 0.4, and 0.8. Generally speaking, these calculated results are consistent with those reported by Ochi *et al* [19]. The band splitting between the lowest and the second lowest conduction bands at *X* point increases with increasing *x*, owing to large spin-orbit coupling associated with Bi ions. The energy difference between the lowest conduction band and the van Hove singularity, that locates along *M*–Γ line, also increases with increasing *x*, because of larger interatomic hopping amplitudes in Bi ions. However, all of calculated band structure exhibits metallic nature, unlike experiments that show insulator-metal-like transition upon Bi doping. We deduce that the contradiction between theory and experiments are due to electron-trapping effect resulted from local structural disorder/instability. All of the crystal structure of the samples presented in this study is assigned to tetragonal *P*4/*nmm* space group using SPXRD measured at room temperature. However, theoretical studies show that the existence of imaginary phonons in the related compounds [10,49-52]. These studies also reported that the total energy of the tetragonal *P*4/*nmm* is not the lowest [51,52]. Indeed, symmetry lowering to monoclinic $P2_1/m$ [53-55] and nanoscale atomic distortion [56-61] have been experimentally observed for $BiCh_2$-based compounds. For another layered compound, $R_2O_2Pn$ with a $Pn^{2-}$ net, insulating nature that originates from lattice distortion has recently been investigated [62-67]. Although both $R_2O_2Bi$ and $R_2O_2Sb$ are expected to be a metallic owing to −2 valence state of Pn, $R_2O_2Sb$ is an insulator, most likely due to the gap arising from the lattice instability [65]. It may be possible that such gap also exists in the present $SbSe_2$-based compounds.

## 4. Conclusion

We synthesized Bi-substituted $SbCh_2$-based compounds $NdO_{0.8}F_{0.2}Sb_{1-x}Bi_xSe_2$ ($x$ = 0–0.8). Crystal structure analysis using SPXRD show that these compounds crystalize in tetragonal *P*4/*nmm* space group at room temperature. Under ambient pressure, insulator-metal-like transition is observed in electrical resistivity at $x$ = 0.5–0.6. For $x$ = 0.3 (Sb-rich composition), a signature of a pressure-induced superconducting transition was observed above 50 GPa. The $T_c^{onset}$ reached 7.65 ± 0.05 K at 70.8 GPa, and the $H_{c2}(0)$ was estimated to be 5.6 T using WHH model. For $x$ = 0.7 (Bi-rich composition), $T_c^{onset}$ = 5.5 ± 0.6 K was observed at 43.5 GPa, which is the almost comparable to that of $x$ = 0.3. However, $H_{c2}$ of $x$ = 0.7 ($H_{c2}$ = 10.9 T) is distinctly higher than that of $x$ = 0.3. The present study demonstrates the possibility of superconductivity in $SbCh_2$-based layered materials using high pressure technique. Furthermore, $NdO_{0.8}F_{0.2}Sb_{1-x}Bi_xSe_2$ are the first report that tunes a wide range of Sb/Bi content ($x$ = 0–0.8) in this family of compounds, and will be a unique platform to explore the difference between electrically insulating $SbCh_2$-based and conducting $BiCh_2$-based systems.

**Acknowledgment**

We thank fruitful discussion with M. Ochi and N. Hirayama. This work was partly supported by JST CREST Grant No. JPMJCR16Q6, JST-Mirai Program Grant Number JPMJMI17A2, JSPS KAKENHI Grant Number 19K15291, 17J05926, 17H05481, 15H05886, and 16H04493, and Iketani Science and Technology Foundation (No. 0301042-A), Japan. A part of the fabrication process of diamond electrodes was supported by NIMS Nanofabrication Platform in Nanotechnology Platform Project sponsored by the Ministry of Education, Culture, Sports, Science and Technology (MEXT), Japan. The part of the high pressure experiments was supported by the Visiting Researcher's Program of Geodynamics Research Center, Ehime University.

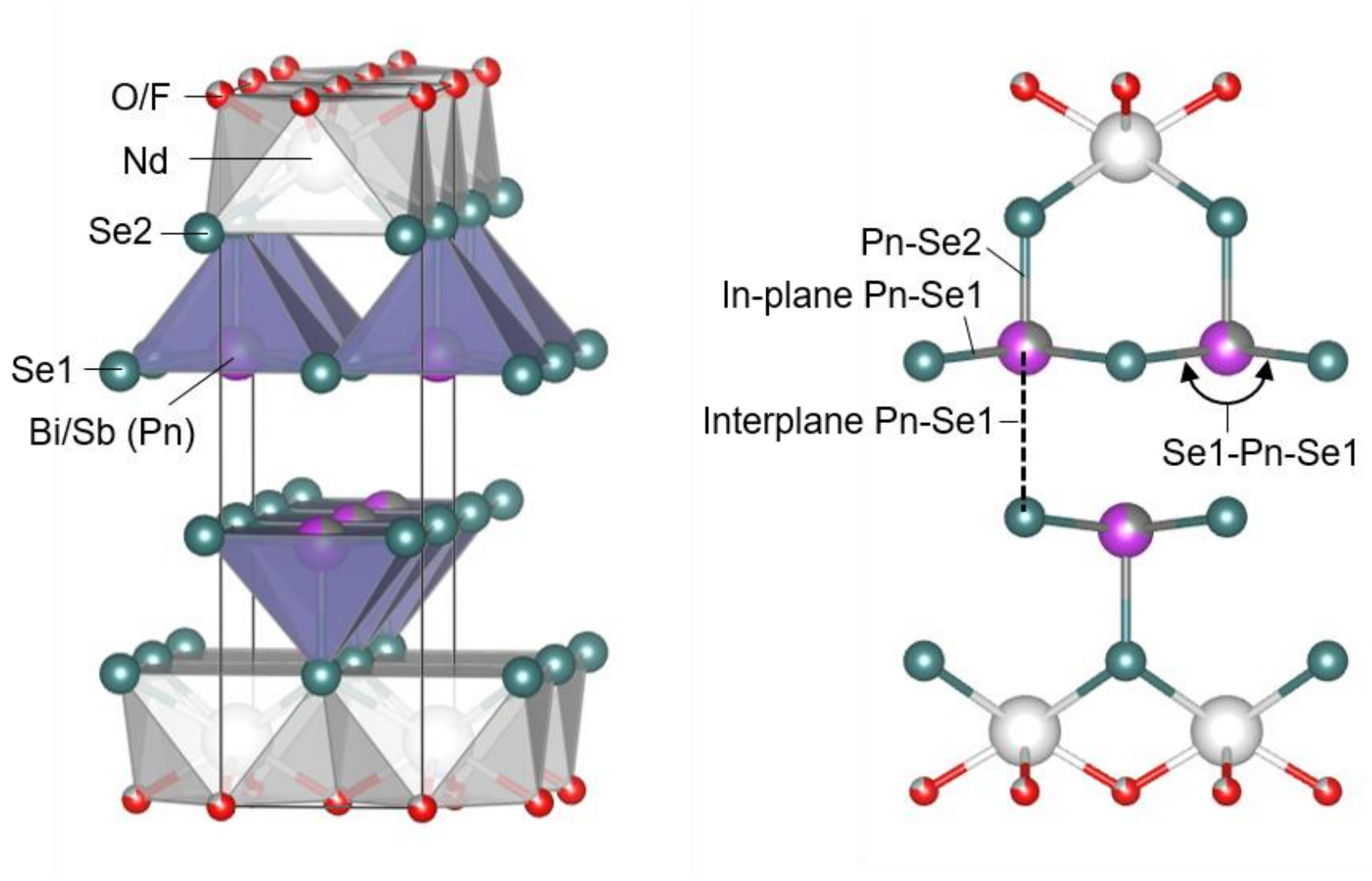

FIG. 1. Crystal structure of $NdO_{0.8}F_{0.2}Sb_{1-x}Bi_xSe_2$ ($x$ = 0.3) which belongs to tetragonal *P*4/*nmm* space group. The black line denotes the unit cell. Se ions have two crystallographic sites: in-plane (Se1) and out-of-plane (Se2). Pnictogen (Sb and Bi) is denoted as Pn.

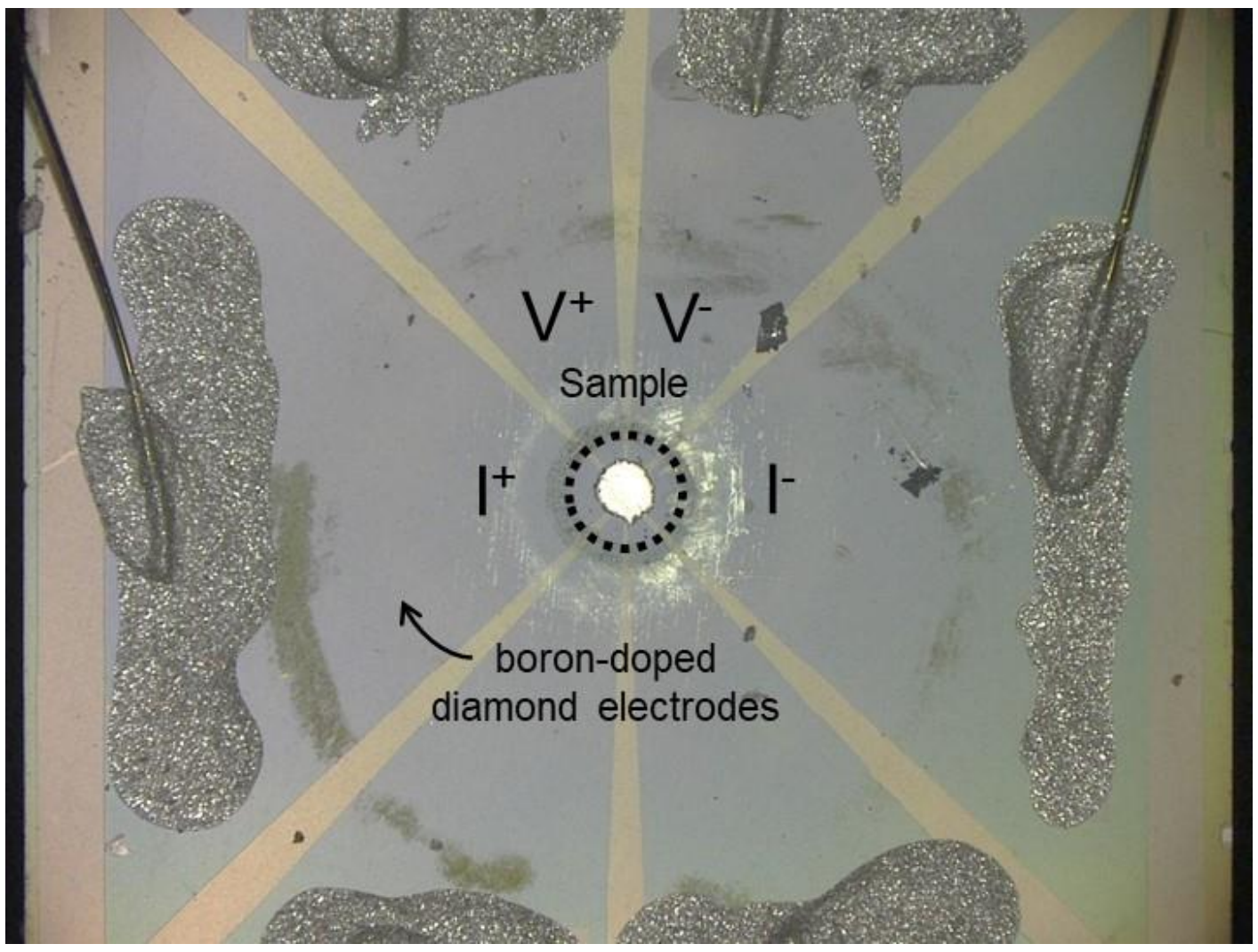

FIG. 2. Optical image of the sample space of DAC with boron-doped diamond electrodes.

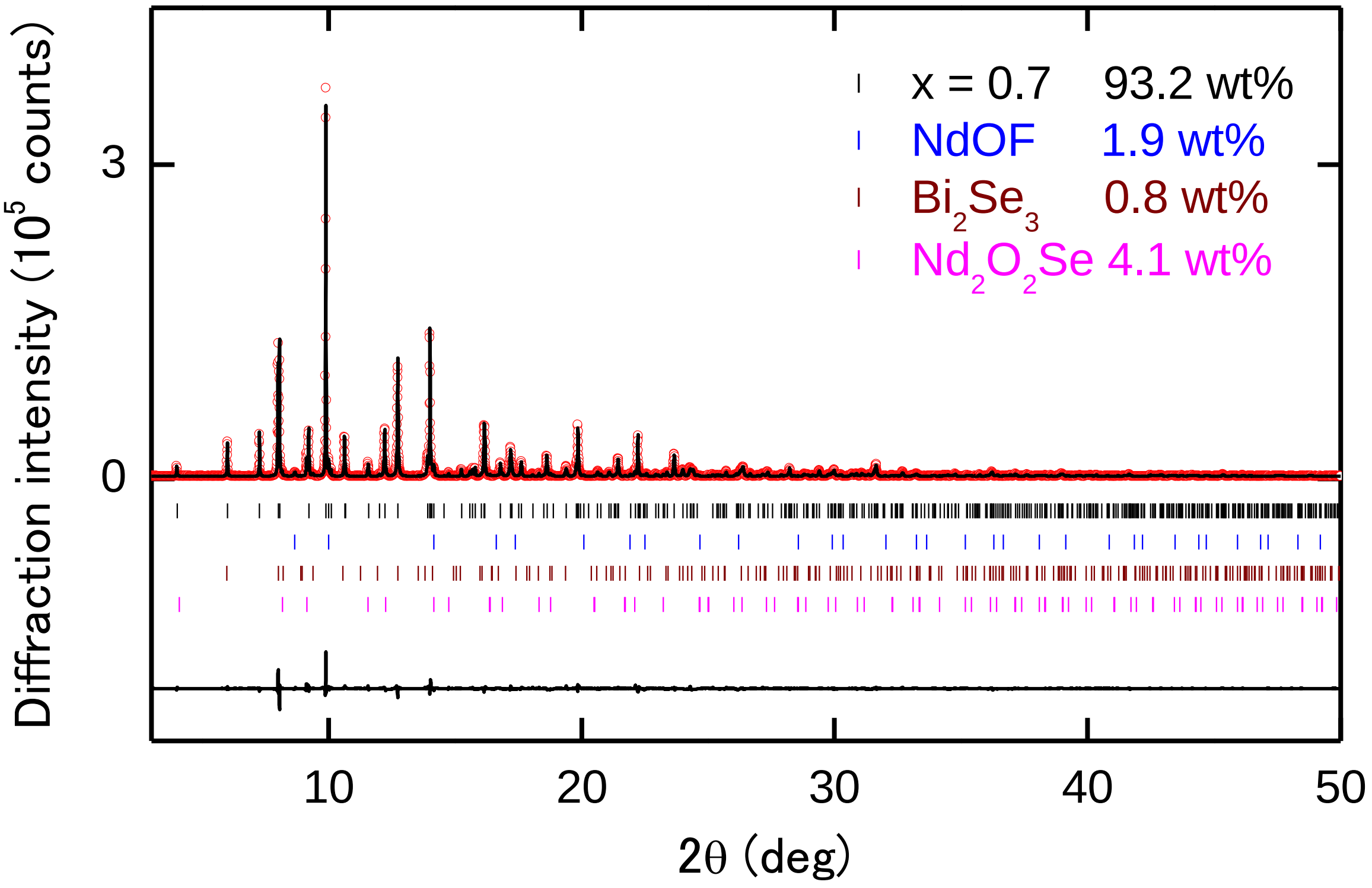

FIG. 3. (a) SPXRD pattern and Rietveld fitting results for $x$ = 0.7 as a representative data of $NdO_{0.8}F_{0.2}Sb_{1-x}Bi_xSe_2$. The measurement was performed at 297 K. The wavelength of the radiation beam was determined to be 0.496345(1) Å. The circles and solid curve represent the observed and calculated patterns, respectively, and the difference between the two is shown at the bottom. The vertical marks indicate the Bragg diffraction positions for $NdO_{0.8}F_{0.2}Sb_{0.3}Bi_{0.7}Se_2$, $Nd_2O_2Se$, NdOF, and $Sb_2Se_3$, from top to bottom, respectively. SPXRD patterns of other samples are shown in Supplemental Material [42].

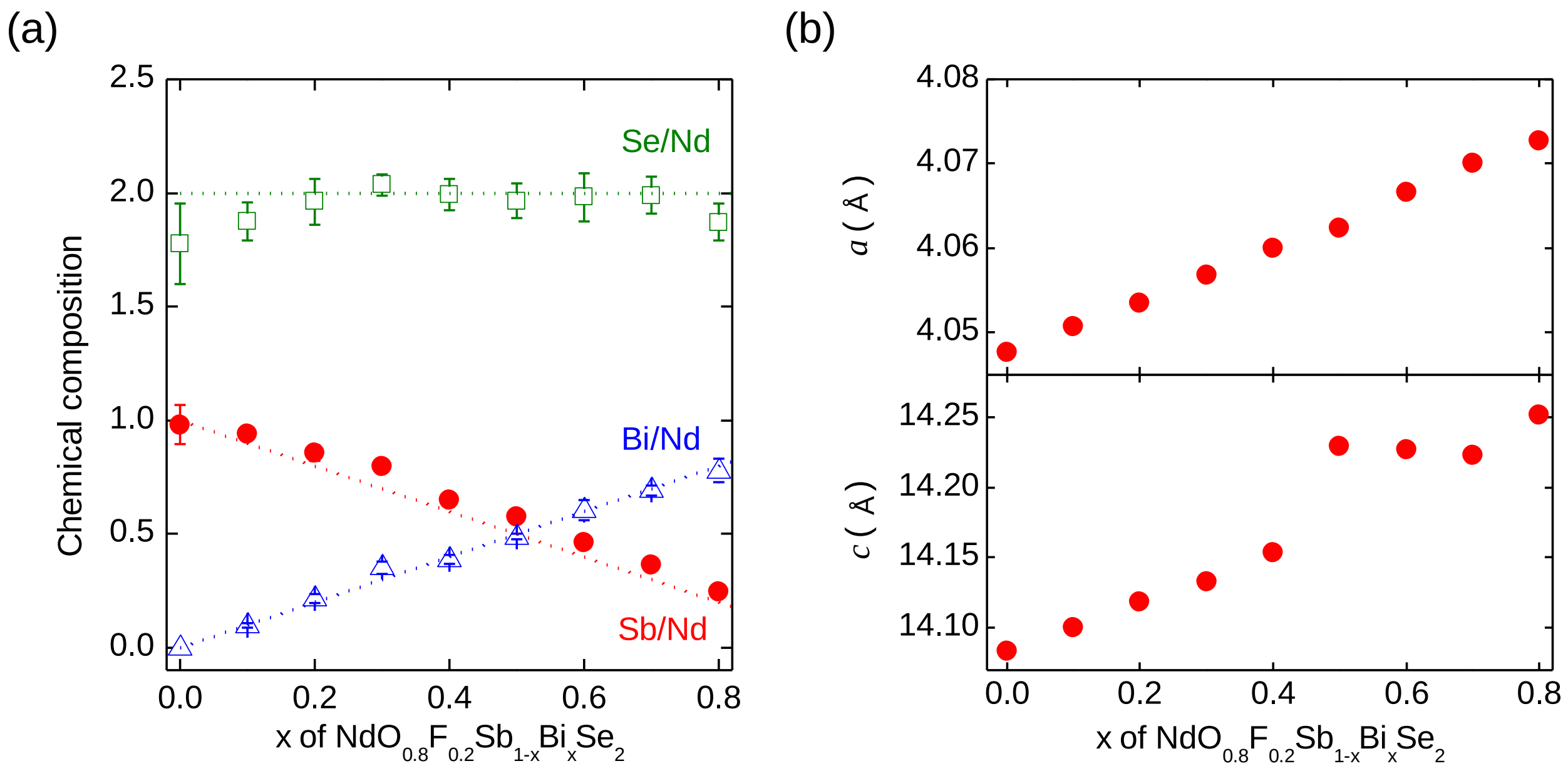

FIG. 4. (a) The *x* dependences of chemical composition ratios of Nd, Sb, Bi, and Se determined using EDX. Dashed lines represent the nominal composition of starting materials. (b) Calculated lattice parameters from Rietveld refinement. The error bars are less than the size of the symbols.

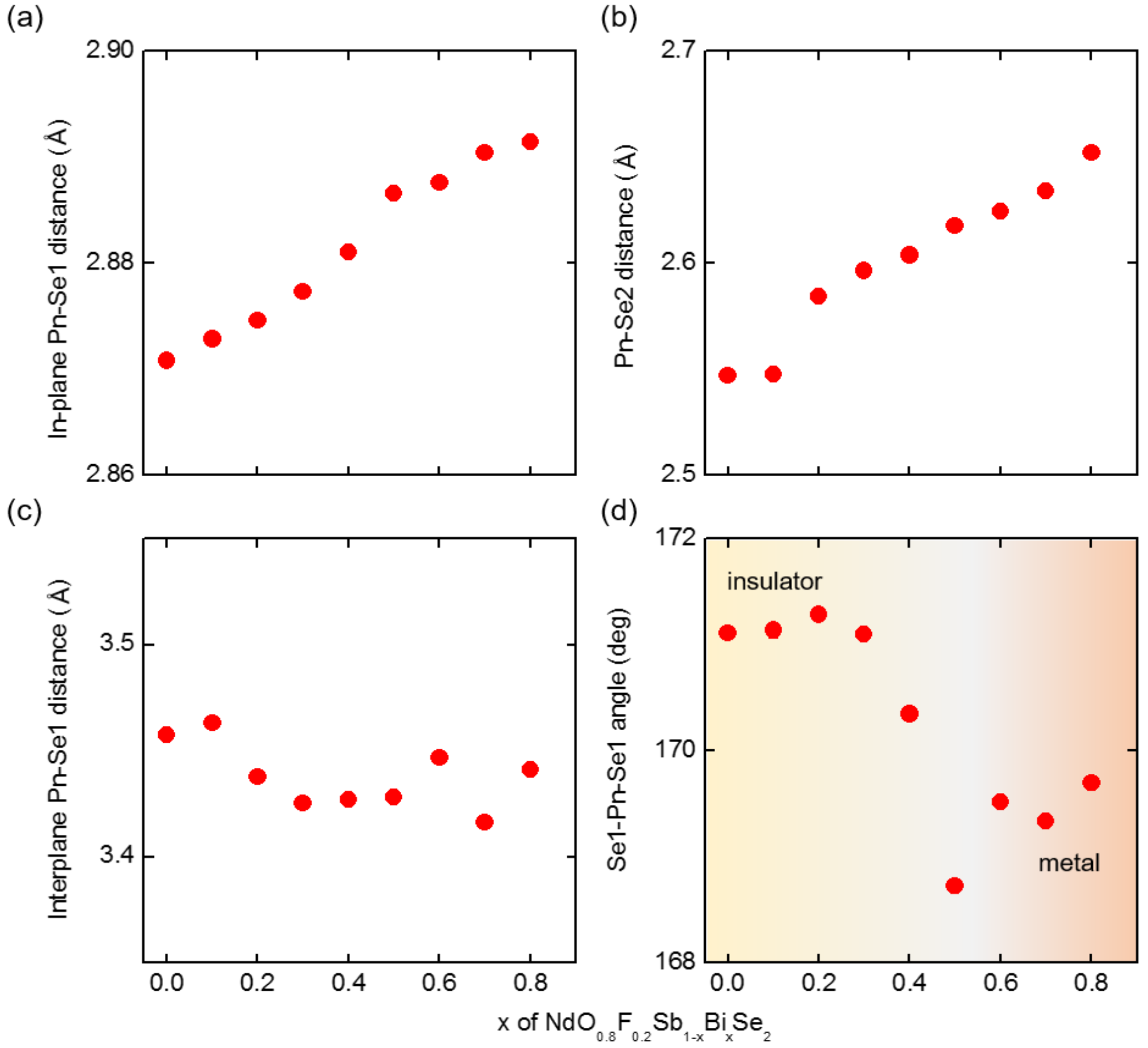

FIG. 5. Selected bond distances and angle for $NdO_{0.8}F_{0.2}Sb_{1-x}Bi_xSe_2$: (a) in-plane Pn-Se1 distance, (b) Pn-Se2 distance, (c) interplane Pn-Se1 distance, and (d) Se1-Pn-Se1 angle. The error bars determined using statistical errors of Rietveld refinement are less than the size of the symbols.

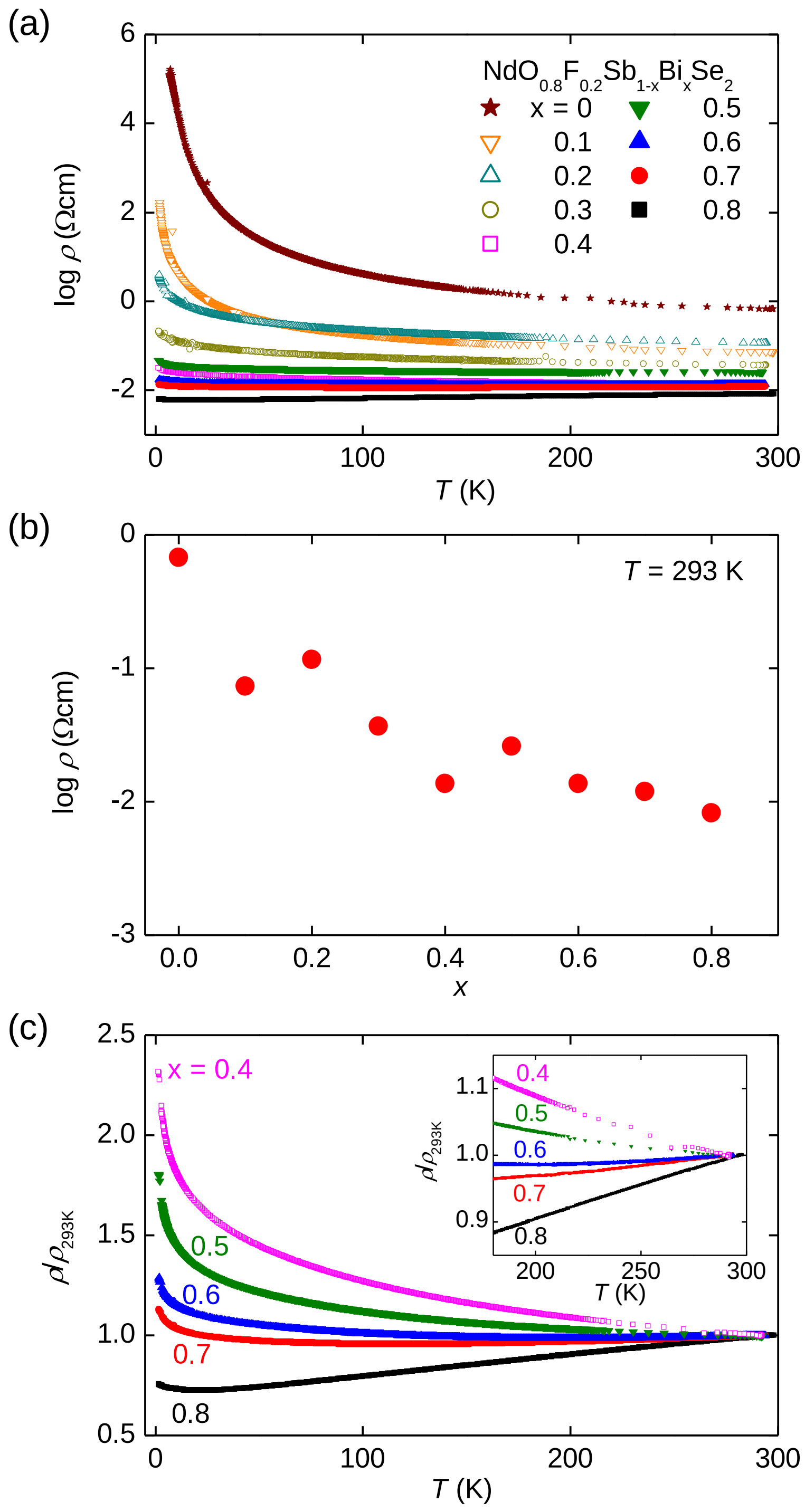

FIG. 6. (a) Temperature ($T$) dependence of logarithmic resistivity ($\rho$) in $NdO_{0.8}F_{0.2}Sb_{1-x}Bi_xSe_2$ ($x$ = 0–0.8) measured at ambient pressure. (b) $\rho$ vs. $T$ at room temperature. (c) Normalized $\rho$ vs. $T$ for $x$ = 0.4–0.8. The inset shows expanded view above 180 K.

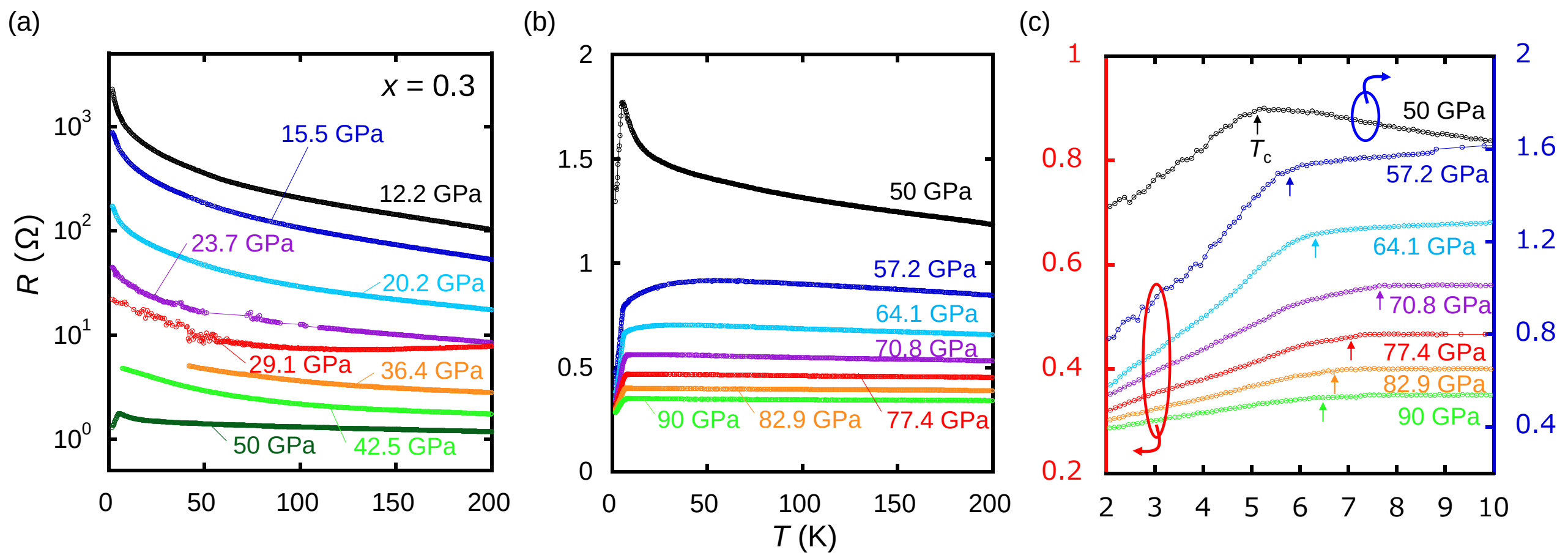

FIG. 7. Temperature dependence of resistance in $x$ = 0.3 under various pressures (a) from 12.2 GPa to 50 GPa, and (b) from 50 GPa to 90 GPa. (c) Temperature dependence of resistance below 10 K under 50 GPa (right axis) and 57.2–90 GPa (left axis). The arrows denote $T_c^{onset}$ (see text for the criterion of $T_c^{onset}$).

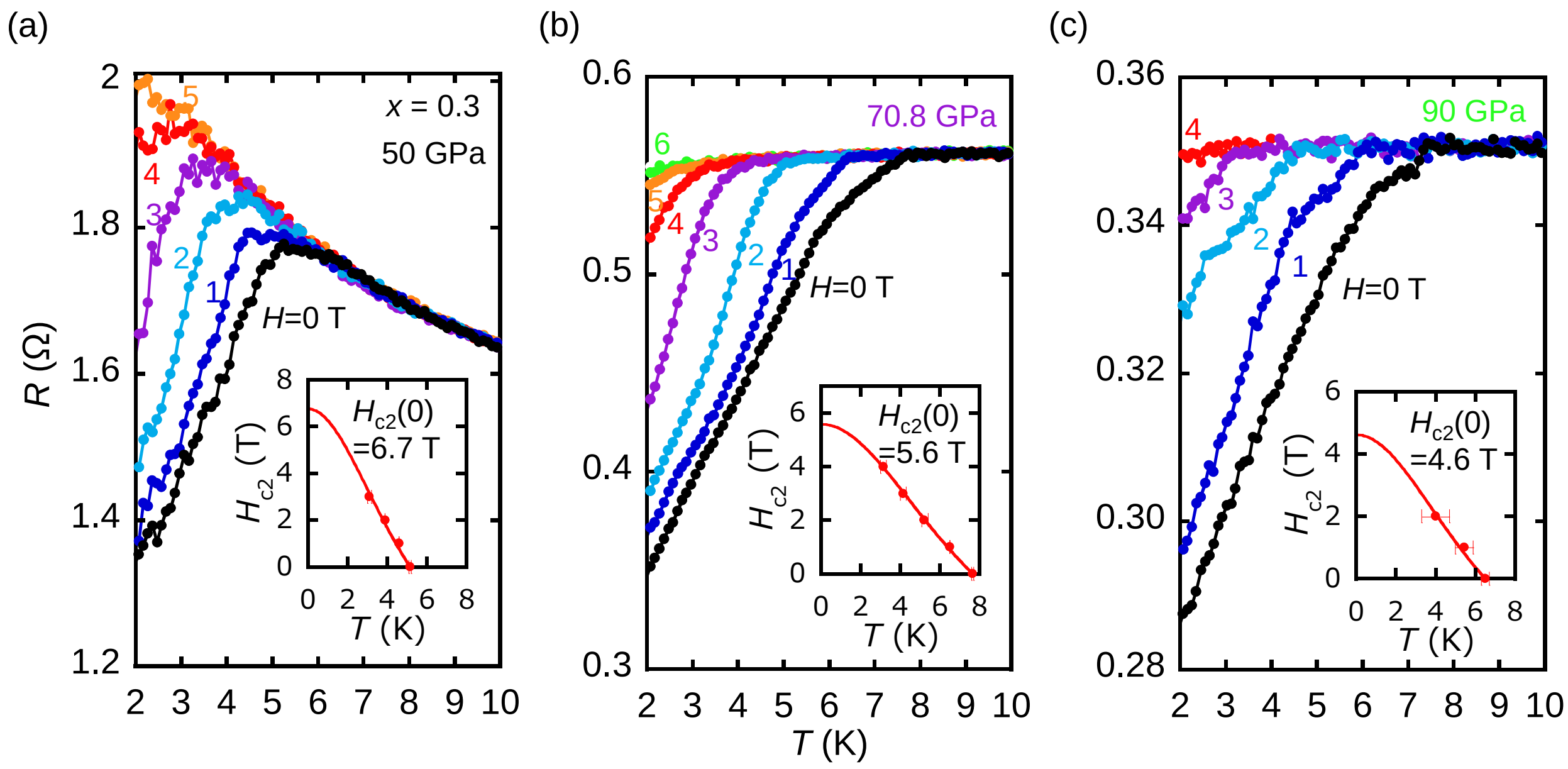

FIG. 8. Temperature dependence of resistance in $x$ = 0.3 in magnetic fields under (a) 50 GPa, (b) 70.8 GPa, and (c) 90 GPa. The insets show magnetic field–temperature phase diagram. Solid line was calculated using WHH model.

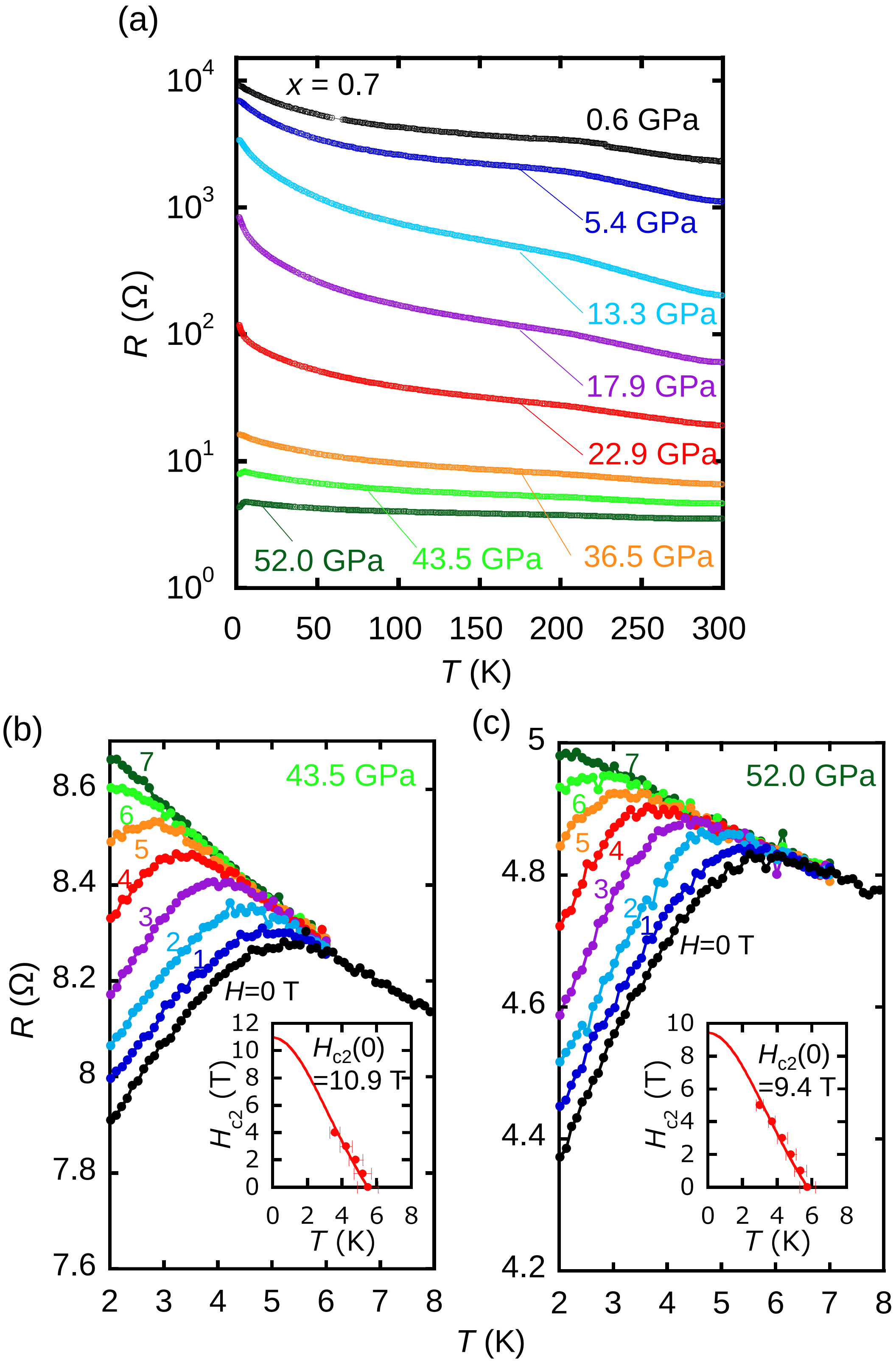

FIG. 9. (a) Temperature dependence of resistance in $x = 0.7$ under various pressure from 0.6 GPa to 52 GPa. (b,c) Temperature dependence of resistance in magnetic fields under (b) 43.5 GPa, and (c) 52 GPa.

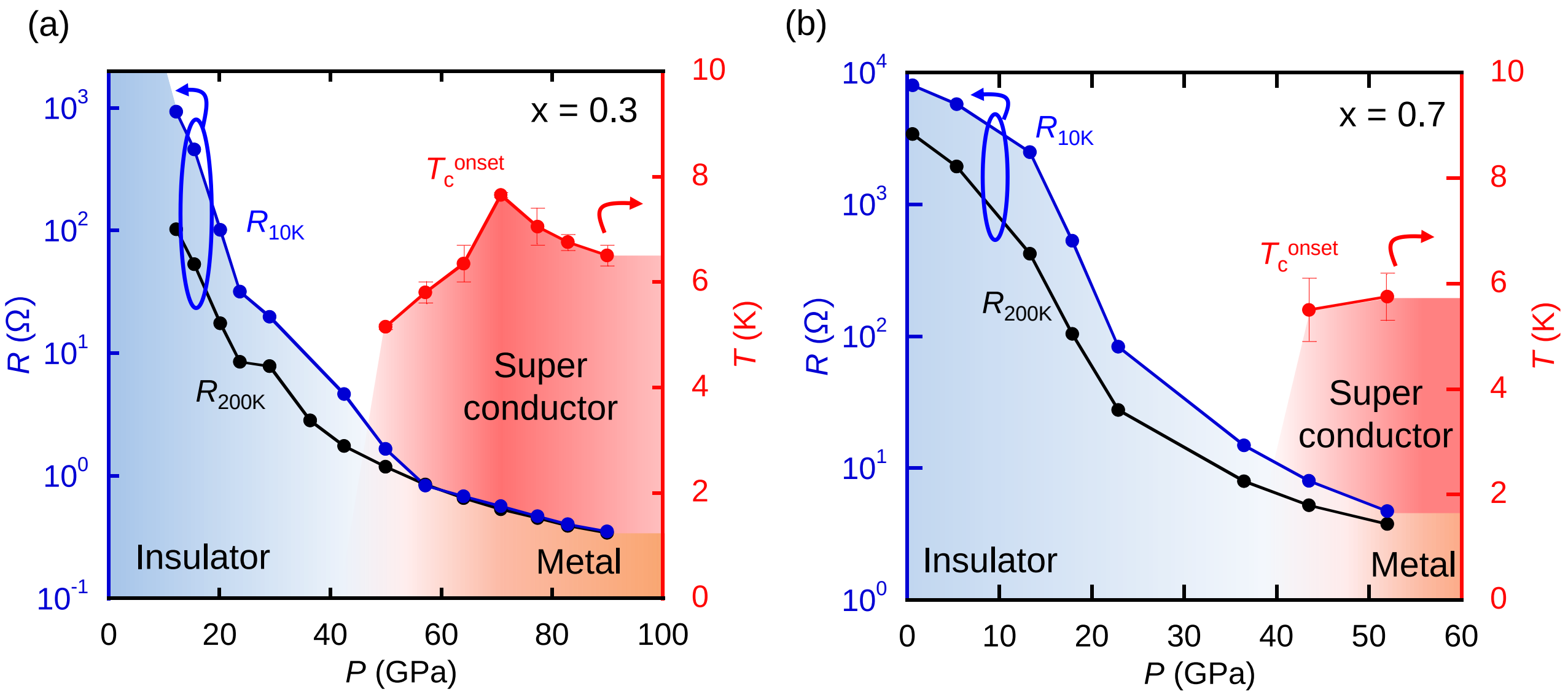

Figure 10. Pressure-phase diagrams for resistance at 200 K and 10 K (left axis) and $T_c^{onset}$ (right axis) for (a) $x$ = 0.3 and (b) $x$ = 0.7.

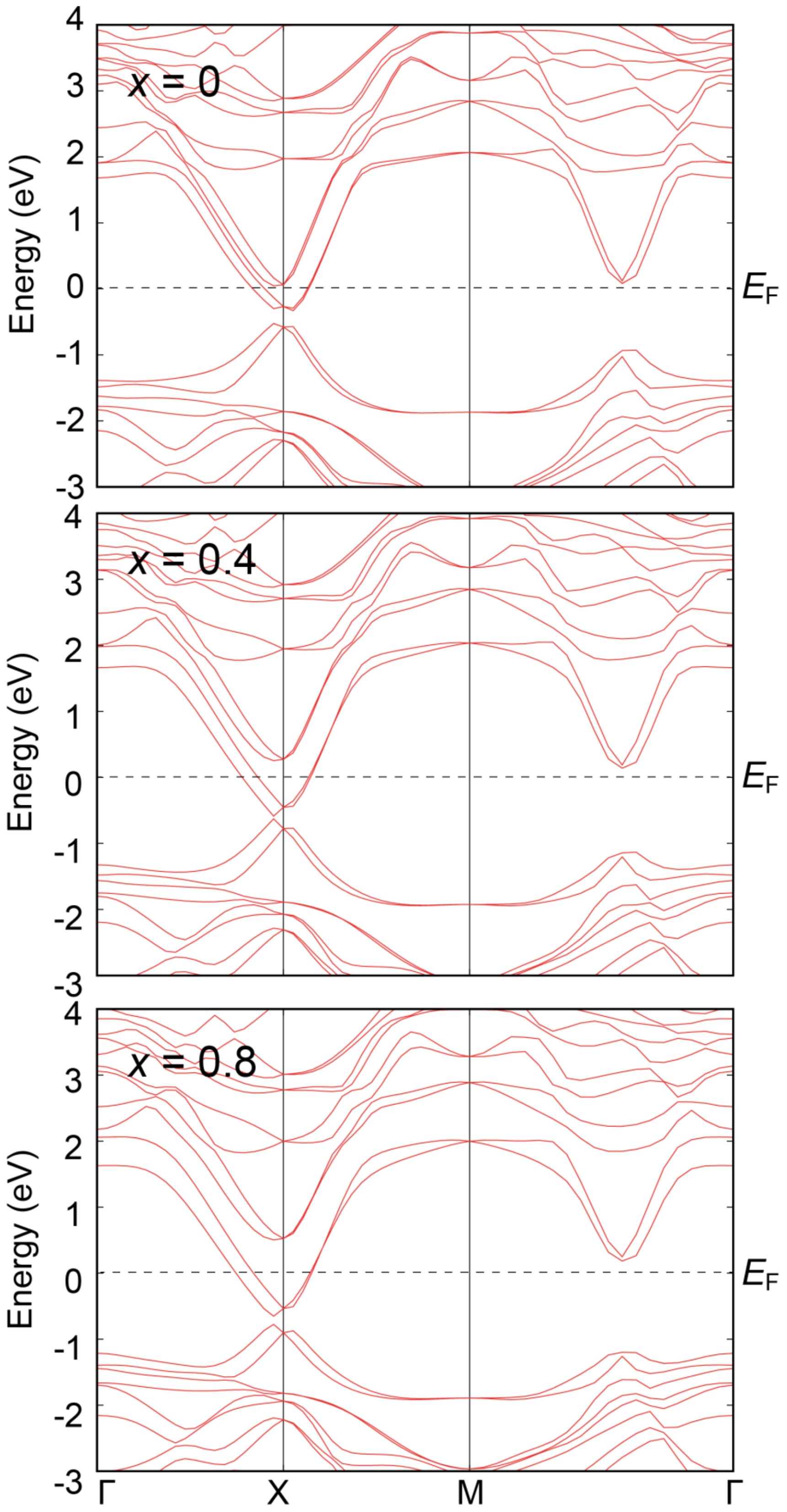

FIG. 11. Band dispersion plot for $NdO_{0.8}F_{0.2}Sb_{1-x}Bi_xSe_2$ ($x$ = 0, 0.4, and 0.8).

Supplemental Materials for

# Pressure-induced superconductivity in layered pnictogen diselenide $NdO_{0.8}F_{0.2}Sb_{1-x}Bi_xSe_2$ ($x$ = 0.3 and 0.7)

Ryo Matsumoto,[1,2] Yosuke Goto,[3,*] Sayaka Yamamoto,[1,2,4] Kenta Sudo,[3,#] Hidetomo Usui,[5] Akira Miura,[6] Chikako Moriyoshi,[7] Yoshihiro Kuroiwa,[7] Shintaro Adachi,[1] Tetsuo Irifune,[8] Hiroyuki Takeya,[1] Hiromi Tanaka,[4] Kazuhiko Kuroki,[9] Yoshikazu Mizuguchi,[3] and Yoshihiko Takano[1,2]

[1]*National Institute for Materials Science, 1-2-1 Sengen, Tsukuba, Ibaraki 305-0047, Japan*

[2]*University of Tsukuba, 1-1-1 Tennodai, Tsukuba, Ibaraki 305-8577, Japan*

[3]*Department of Physics, Tokyo Metropolitan University, 1-1 Minami-osawa, Hachioji, Tokyo 192-0397, Japan*

[4]*National Institute of Technology, Yonago College, 4448 Hikona, Yonago, Tottori 683-8502, Japan*

[5]*Department of Physics and Materials Science, Shimane University, Matsue, 690-8504, Japan*

[6] *Faculty of Engineering, Hokkaido University, Kita 13, Nishi 8 Sapporo 060-8628, Japan*

[7] *Department of Physical Science, Hiroshima University, 1-3-1 Kagamiyama, Higashihiroshima, Hiroshima 739-8526, Japan*

[8]*Geodynamics Research Center, Ehime University, Matsuyama, Ehime 790-8577, Japan*

[9]*Department of Physics, Osaka University, Toyonaka, Osaka 560-0043, Japan*

[#]*Present address: High Field Laboratory for Superconducting Materials, Institute for Materials Research, Tohoku University, Sendai 980-8577, Japan*

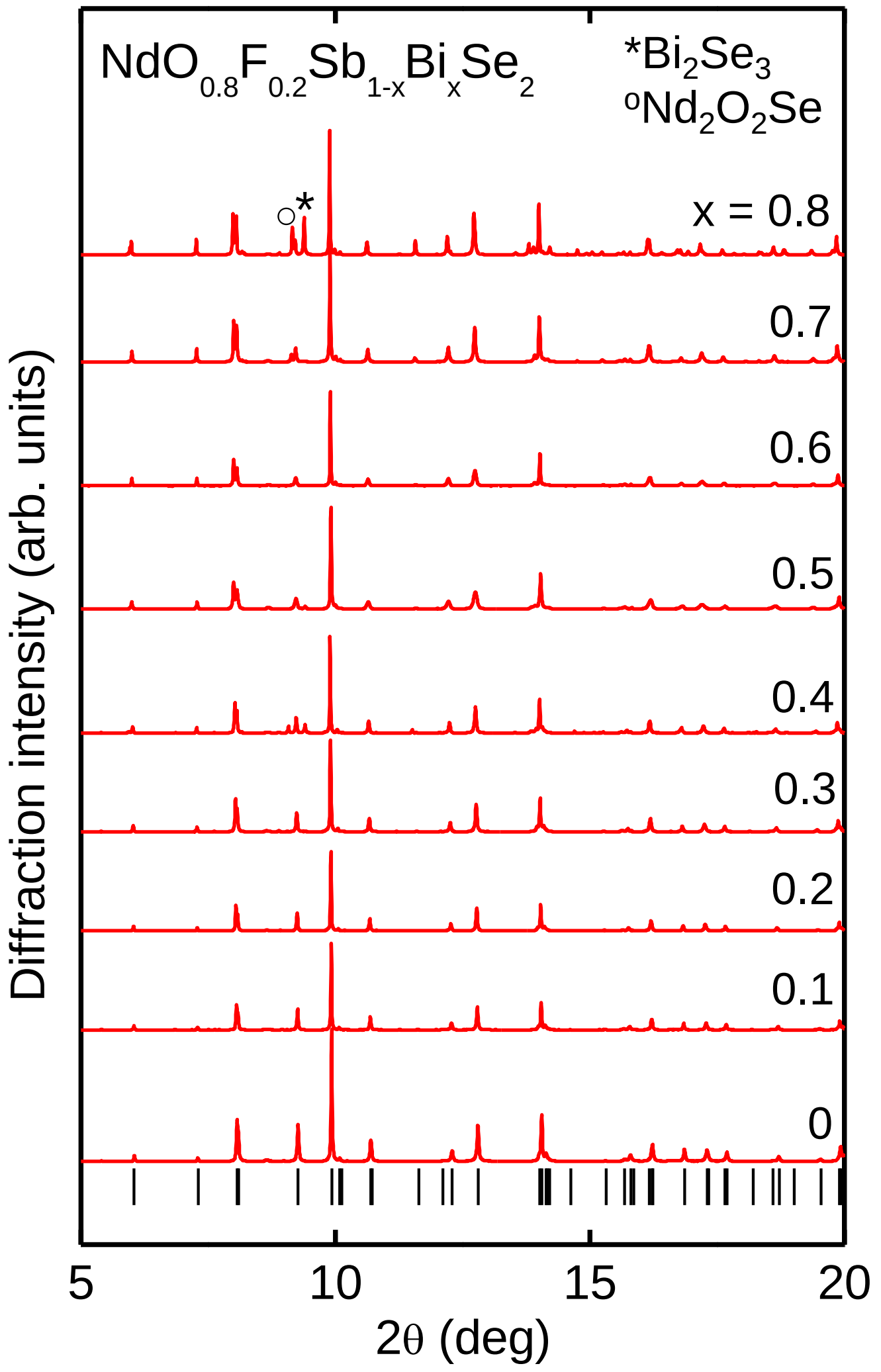

FIG. S1. SPXRD patterns of $NdO_{0.8}F_{0.2}Sb_{1-x}Bi_xSe_2$. The vertical marks at the bottom show the Bragg diffraction angles of $NdO_{0.8}F_{0.2}SbSe_2$. The asterisk and open circle represent the diffraction peaks due to $Bi_2Se_3$ and $Nd_2O_2Se$, respectively.

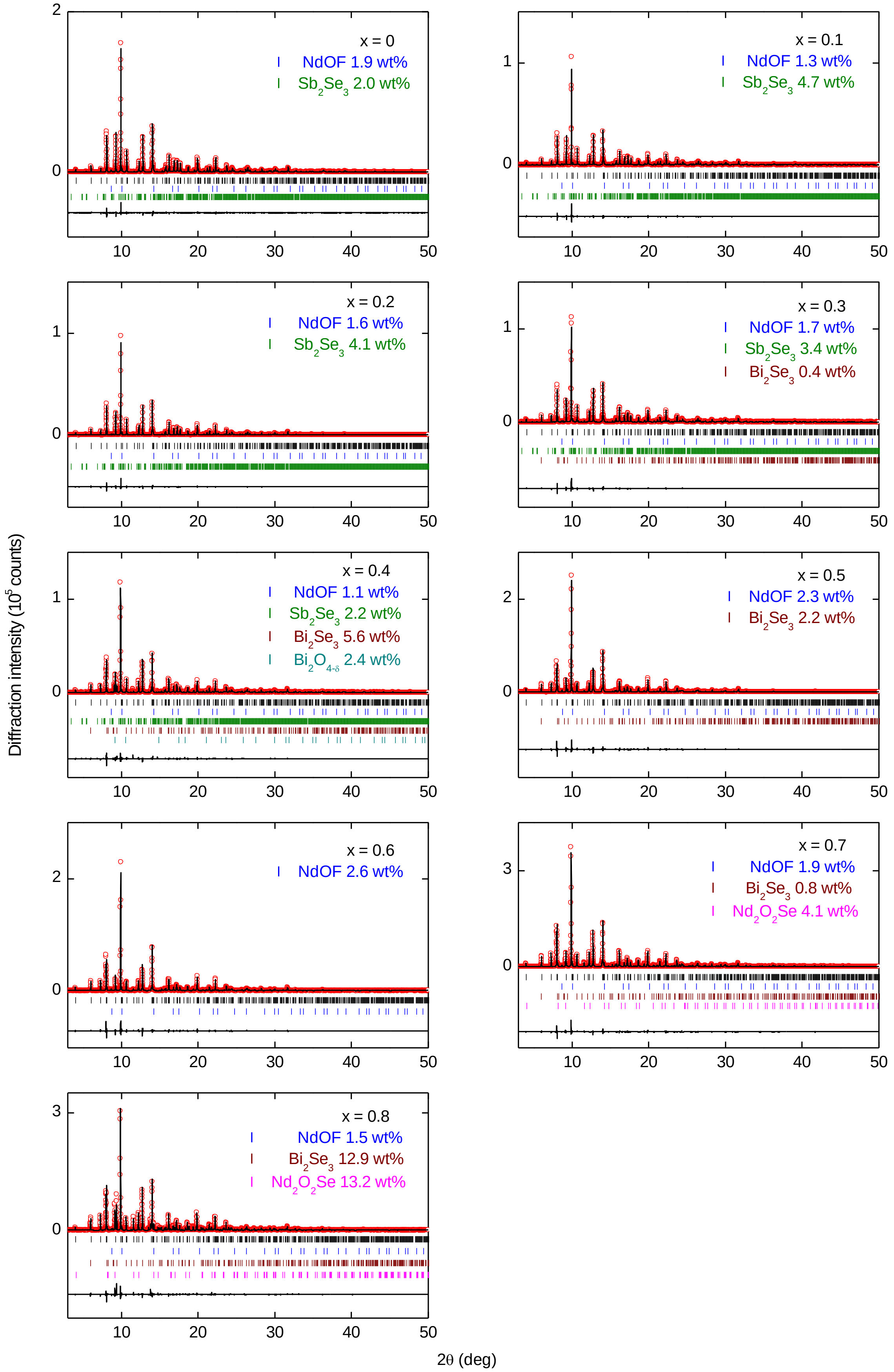

FIG. S2. SPXRD patterns and the results of Rietveld refinement for $NdO_{0.8}F_{0.2}Sb_{1-x}Bi_xSe_2$.

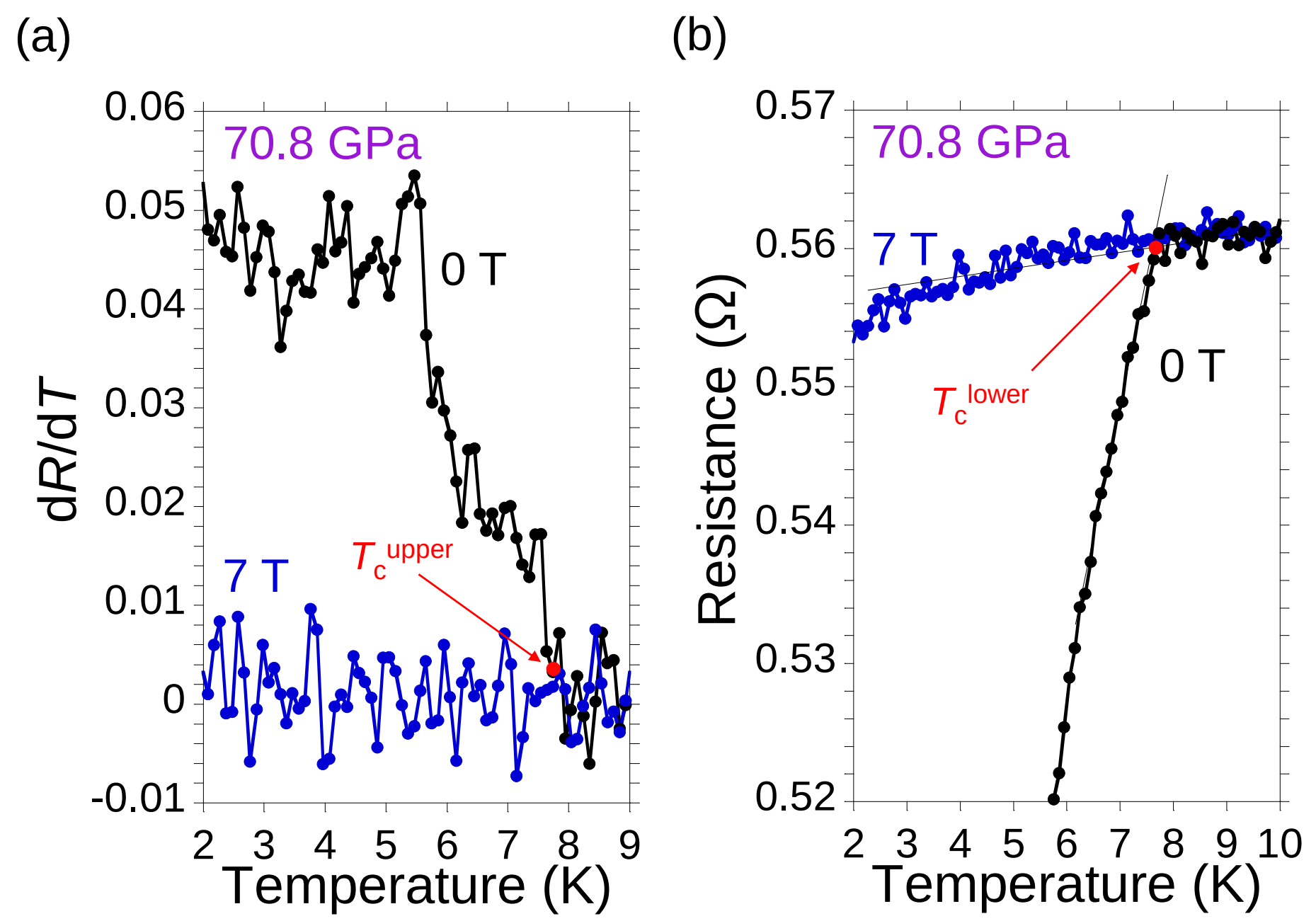

FIG. S3. Temperature dependence of resistance for $x$ = 0.3 at 70.8 GPa. (a) The differential curve under 0 T and 7 T. $T_c^{upper}$ was determined from a branch point of these two curves. (b) $R$ vs. $T$. $T_c^{lower}$ was determined from an intersection point between the strait line of the normal resistance region and extended line from resistance after superconducting transition under 0 T and 7 T

**Table S1.** Crystal structure parameters and reliability factors of $NdO_{0.8}F_{0.2}Sb_{1-x}Bi_xSe_2$ obtained by Rietveld analysis. Values in parentheses are standard deviations in the last digits. For Sb and Se(1) sites, anisotropic displacement parameters are refined. The data for $x = 0$ is taken from literature (Y. Goto *et al.* J. Phys. Soc. Jpn. 88, 024705 (2019).).

| | | | | | $x = 0$ | |
|---|---|---|---|---|---|---|
| Lattice system | | | | | Tetragonal | |
| Space group | | | | | *P*4/*nmm* (No. 129) | |
| Lattice parameters | | | | | $a$ = 4.04769(1) Å, $c$ = 14.08331(9) Å | |
| atom | site | $g$ | $x$ | $y$ | $z$ | $U$ (Å$^2$) |
| Nd | 2*c* | 1.0 | 0 | 1/2 | 0.08684(4) | $U_{iso}$ = 0.00813(15) |
| O | 2*a* | 0.8 | 0 | 0 | 0 | $U_{iso}$ = 0.0118(18) |
| F | 2*a* | 0.2 | 0 | 0 | 0 | $U_{iso}$ = 0.0118(18) |
| Sb | 2*c* | 1.0 | 0 | 1/2 | 0.63066(5) | $U_{11}$ = 0.0244(4) |
| | | | | | | $U_{33}$ = 0.0098(5) |
| | | | | | | $U_{eq}$ = 0.020 |
| Se(1) | 2*c* | 1.0 | 0 | 1/2 | 0.38514(8) | $U_{11}$ = 0.0309(6) |
| | | | | | | $U_{33}$ = 0.0230(9) |
| | | | | | | $U_{eq}$ = 0.028 |
| Se(2) | 2*c* | 1.0 | 0 | 1/2 | 0.81153(7) | $U_{iso}$ = 0.0096(3) |
| $R_{wp}$ | | | | | 5.69% | |
| $R_B$ | | | | | 1.15% | |
| GOF | | | | | 2.40 | |

| $x = 0.1$ | | | | | | |
|---|---|---|---|---|---|---|
| Lattice system | Tetragonal | | | | | |
| Space group | *P*4/*nmm* (No. 129) | | | | | |
| Lattice parameters | $a$ = 4.05067(1) Å, $c$ = 14.10060(8) Å | | | | | |
| atom | site | $g$ | $x$ | $y$ | $z$ | $U$ (Å$^2$) |
| Nd | 2*c* | 1.0 | 0 | 1/2 | 0.08689(5) | $U_{iso}$ = 0.00893(18) |
| O | 2*a* | 0.8 | 0 | 0 | 0 | $U_{iso}$ = 0.019(2) |
| F | 2*a* | 0.2 | 0 | 0 | 0 | $U_{iso}$ = 0.019(2) |
| Sb | 2*c* | 0.9 | 0 | 1/2 | 0.62990(6) | $U_{11}$ = 0.0257(3) |
| | | | | | | $U_{33}$ = 0.0069(5) |
| | | | | | | $U_{eq}$ = 0.019 |
| Bi | 2*c* | 0.1 | 0 | 1/2 | 0.62990(6) | $U_{11}$ = 0.0257(3) |
| | | | | | | $U_{33}$ = 0.0069(5) |
| | | | | | | $U_{eq}$ = 0.019 |
| Se(1) | 2*c* | 1.0 | 0 | 1/2 | 0.38546(10) | $U_{11}$ = 0.0315(6) |
| | | | | | | $U_{33}$ = 0.0173(10) |
| | | | | | | $U_{eq}$ = 0.026 |
| Se(2) | 2*c* | 1.0 | 0 | 1/2 | 0.81214(9) | $U_{iso}$ = 0.0100(3) |
| $R_{wp}$ | 6.63% | | | | | |
| $R_B$ | 1.77% | | | | | |
| GOF | 2.06 | | | | | |
| | | | $x$ = 0.2 | | | |
| Lattice system | | | Tetragonal | | | |
| Space group | | | $P4/nmm$ (No. 129) | | | |
| Lattice parameters | | | $a$ = 4.05347(1) Å, $c$ = 14.11837(8) Å | | | |
| atom | site | $g$ | $x$ | $y$ | $z$ | $U$ (Å$^2$) |
| Nd | 2$c$ | 1.0 | 0 | 1/2 | 0.08635(5) | $U_{iso}$ = 0.00872(18) |
| O | 2$a$ | 0.8 | 0 | 0 | 0 | $U_{iso}$ = 0.020(2) |
| F | 2$a$ | 0.2 | 0 | 0 | 0 | $U_{iso}$ = 0.020(2) |
| Sb | 2$c$ | 0.8 | 0 | 1/2 | 0.62949(5) | $U_{11}$ = 0.0236(3) |
| | | | | | | $U_{33}$ = 0.0102(5) |
| | | | | | | $U_{eq}$ = 0.019 |
| Bi | 2$c$ | 0.2 | 0 | 1/2 | 0.62949(5) | $U_{11}$ = 0.0236(3) |
| | | | | | | $U_{33}$ = 0.0102(5) |
| | | | | | | $U_{eq}$ = 0.019 |
| Se(1) | 2$c$ | 1.0 | 0 | 1/2 | 0.38614(9) | $U_{11}$ = 0.0271(6) |
| | | | | | | $U_{33}$ = 0.0139(9) |
| | | | | | | $U_{eq}$ = 0.022 |
| Se(2) | 2$c$ | 1.0 | 0 | 1/2 | 0.81266(8) | $U_{iso}$ = 0.0094(3) |
| $R_{wp}$ | | | 5.53% | | | |
| $R_B$ | | | 2.08% | | | |
| GOF | | | 1.80 | | | |
| | | | $x$ = 0.3 | | | |
| Lattice system | | | Tetragonal | | | |
| Space group | | | $P4/nmm$ (No. 129) | | | |
| Lattice parameters | | | $a$ = 4.05684(2) Å, $c$ = 14.13303(9) Å | | | |
| atom | site | $g$ | $x$ | $y$ | $z$ | $U$ (Å$^2$) |
| Nd | 2$c$ | 1.0 | 0 | 1/2 | 0.08615(5) | $U_{iso}$ = 0.0084(2) |
| O | 2$a$ | 0.8 | 0 | 0 | 0 | $U_{iso}$ = 0.032(3) |
| F | 2$a$ | 0.2 | 0 | 0 | 0 | $U_{iso}$ = 0.032(3) |
| Sb | 2$c$ | 0.7 | 0 | 1/2 | 0.62903(5) | $U_{11}$ = 0.0223(3) |
| | | | | | | $U_{33}$ = 0.0090(5) |
| | | | | | | $U_{eq}$ = 0.018 |
| Bi | 2$c$ | 0.3 | 0 | 1/2 | 0.62903(5) | $U_{11}$ = 0.0223(3) |
| | | | | | | $U_{33}$ = 0.0090(5) |
| | | | | | | $U_{eq}$ = 0.018 |
| Se(1) | 2$c$ | 1.0 | 0 | 1/2 | 0.38681(9) | $U_{11}$ = 0.0226(6) |
| | | | | | | $U_{33}$ = 0.0162(10) |
| | | | | | | $U_{eq}$ = 0.020 |
| Se(2) | 2$c$ | 1.0 | 0 | 1/2 | 0.81299(9) | $U_{iso}$ = 0.0095(3) |
| $R_{wp}$ | | | 5.76% | | | |
| $R_B$ | | | 1.48% | | | |
| GOF | | | 2.21 | | | |

| $x = 0.4$ | | | | | | |
|---|---|---|---|---|---|---|
| Lattice system | | | Tetragonal | | | |
| Space group | | | *P*4/*nmm* (No. 129) | | | |
| Lattice parameters | | | $a$ = 4.05999(2) Å, $c$ = 14.15324(15) Å | | | |
| atom | site | $g$ | $x$ | $y$ | $z$ | $U$ (Å$^2$) |
| Nd | 2*c* | 1.0 | 0 | 1/2 | 0.08559(8) | $U_{iso}$ = 0.0078(3) |
| O | 2*a* | 0.8 | 0 | 0 | 0 | $U_{iso}$ = 0.031(5) |
| F | 2*a* | 0.2 | 0 | 0 | 0 | $U_{iso}$ = 0.031(5) |
| Sb | 2*c* | 0.6 | 0 | 1/2 | 0.62941(8) | $U_{11}$ = 0.0247(5) |
| | | | | | | $U_{33}$ = 0.0064(8) |
| | | | | | | $U_{eq}$ = 0.019 |
| Bi | 2*c* | 0.4 | 0 | 1/2 | 0.62941(8) | $U_{11}$ = 0.0247(5) |
| | | | | | | $U_{33}$ = 0.0064(8) |
| | | | | | | $U_{eq}$ = 0.019 |
| Se(1) | 2*c* | 1.0 | 0 | 1/2 | 0.38749(15) | $U_{11}$ = 0.0206(9) |
| | | | | | | $U_{33}$ = 0.0077(15) |
| | | | | | | $U_{eq}$ = 0.016 |
| Se(2) | 2*c* | 1.0 | 0 | 1/2 | 0.81375(15) | $U_{iso}$ = 0.0096(5) |
| $R_{wp}$ | | | 9.18% | | | |
| $R_B$ | | | 2.13% | | | |
| GOF | | | 3.37 | | | |

| $x$ = 0.5 | | | | | | |
|---|---|---|---|---|---|---|
| Lattice system | | | | Tetragonal | | |
| Space group | | | | $P4/nmm$ (No. 129) | | |
| Lattice parameters | | | | $a$ = 4.06245(3) Å, $c$ = 14.2296(2) Å | | |
| atom | site | $g$ | $x$ | $y$ | $z$ | $U$ (Å$^2$) |
| Nd | 2$c$ | 1.0 | 0 | 1/2 | 0.08490(8) | $U_{iso}$ = 0.0070(3) |
| O | 2$a$ | 0.8 | 0 | 0 | 0 | $U_{iso}$ = 0.021(4) |
| F | 2$a$ | 0.2 | 0 | 0 | 0 | $U_{iso}$ = 0.021(4) |
| Sb | 2$c$ | 0.5 | 0 | 1/2 | 0.63039(8) | $U_{11}$ = 0.0219(4) |
| | | | | | | $U_{33}$ = 0.0003(8) |
| | | | | | | $U_{eq}$ = 0.015 |
| Bi | 2$c$ | 0.5 | 0 | 1/2 | 0.63039(8) | $U_{11}$ = 0.0219(4) |
| | | | | | | $U_{33}$ = 0.0003(8) |
| | | | | | | $U_{eq}$ = 0.015 |
| Se(1) | 2$c$ | 1.0 | 0 | 1/2 | 0.38956(15) | $U_{11}$ = 0.0182(9) |
| | | | | | | $U_{33}$ = 0.0092(17) |
| | | | | | | $U_{eq}$ = 0.015 |
| Se(2) | 2$c$ | 1.0 | 0 | 1/2 | 0.81431(15) | $U_{iso}$ = 0.0106(6) |
| $R_{wp}$ | | | | 7.97% | | |
| $R_B$ | | | | 1.90% | | |
| GOF | | | | 4.68 | | |

| | $x = 0.6$ | | | | | |
|---|---|---|---|---|---|---|
| Lattice system | Tetragonal | | | | | |
| Space group | *P*4/*nmm* (No. 129) | | | | | |
| Lattice parameters | $a$ = 4.06658(2) Å, $c$ = 14.22685(18) Å | | | | | |
| atom | site | $g$ | $x$ | $y$ | $z$ | $U$ (Å$^2$) |
| Nd | 2*c* | 1.0 | 0 | 1/2 | 0.08497(9) | $U_{iso}$ = 0.0119(4) |
| O | 2*a* | 0.8 | 0 | 0 | 0 | $U_{iso}$ = 0.01 |
| F | 2*a* | 0.2 | 0 | 0 | 0 | $U_{iso}$ = 0.01 |
| Sb | 2*c* | 0.4 | 0 | 1/2 | 0.63036(9) | $U_{11}$ = 0.0164(4) |
| | | | | | | $U_{33}$ = 0.0003 |
| | | | | | | $U_{eq}$ = 0.011 |
| Bi | 2*c* | 0.6 | 0 | 1/2 | 0.63036(9) | $U_{11}$ = 0.0164(4) |
| | | | | | | $U_{33}$ = 0.0003 |
| | | | | | | $U_{eq}$ = 0.011 |
| Se(1) | 2*c* | 1.0 | 0 | 1/2 | 0.38815(16) | $U_{11}$ = 0.0265(10) |
| | | | | | | $U_{33}$ = 0.0044(17) |
| | | | | | | $U_{eq}$ = 0.019 |
| Se(2) | 2*c* | 1.0 | 0 | 1/2 | 0.81477(17) | $U_{iso}$ = 0.0145(6) |
| $R_{wp}$ | 8.48% | | | | | |
| $R_B$ | 2.35% | | | | | |
| GOF | 4.28 | | | | | |

| $x = 0.7$ | | | | | | |
| --- | --- | --- | --- | --- | --- | --- |
| Lattice system | Tetragonal | | | | | |
| Space group | *P*4/*nmm* (No. 129) | | | | | |
| Lattice parameters | $a$ = 4.07004(2) Å, $c$ = 14.22277(13) Å | | | | | |
| atom | site | $g$ | $x$ | $y$ | $z$ | $U$ (Å$^2$) |
| Nd | 2*c* | 1.0 | 0 | 1/2 | 0.08546(7) | $U_{iso}$ = 0.0127(3) |
| O | 2*a* | 0.8 | 0 | 0 | 0 | $U_{iso}$ = 0.004(2) |
| F | 2*a* | 0.2 | 0 | 0 | 0 | $U_{iso}$ = 0.004(2) |
| Sb | 2*c* | 0.3 | 0 | 1/2 | 0.62956(6) | $U_{11}$ = 0.0189(3) |
| | | | | | | $U_{33}$ = 0.0095(5) |
| | | | | | | $U_{eq}$ = 0.016 |
| Bi | 2*c* | 0.7 | 0 | 1/2 | 0.62956(6) | $U_{11}$ = 0.0189(3) |
| | | | | | | $U_{33}$ = 0.0095(5) |
| | | | | | | $U_{eq}$ = 0.016 |
| Se(1) | 2*c* | 1.0 | 0 | 1/2 | 0.38913(13) | $U_{11}$ = 0.0215(7) |
| | | | | | | $U_{33}$ = 0.0170(14) |
| | | | | | | $U_{eq}$ = 0.020 |
| Se(2) | 2*c* | 1.0 | 0 | 1/2 | 0.81437(12) | $U_{iso}$ = 0.0141(4) |
| $R_{wp}$ | 7.10% | | | | | |
| $R_B$ | 2.04% | | | | | |
| GOF | 4.89 | | | | | |

| $x = 0.8$ | | | | | | |
|---|---|---|---|---|---|---|
| Lattice system | | | | Tetragonal | | |
| Space group | | | | *P*4/*nmm* (No. 129) | | |
| Lattice parameters | | | | $a$ = 4.07266(2) Å, $c$ = 14.25275(14) Å | | |
| atom | site | *g* | *x* | *y* | *z* | *U* (Å$^2$) |
| Nd | 2*c* | 1.0 | 0 | 1/2 | 0.08434(10) | $U_{iso}$ = 0.0116(4) |
| O | 2*a* | 0.8 | 0 | 0 | 0 | $U_{iso}$ = 0.013(4) |
| F | 2*a* | 0.2 | 0 | 0 | 0 | $U_{iso}$ = 0.013(4) |
| Sb | 2*c* | 0.2 | 0 | 1/2 | 0.62986(8) | $U_{11}$ = 0.0182(3) |
| | | | | | | $U_{33}$ = 0.0139(7) |
| | | | | | | $U_{eq}$ = 0.017 |
| Bi | 2*c* | 0.8 | 0 | 1/2 | 0.62986(8) | $U_{11}$ = 0.0182(3) |
| | | | | | | $U_{33}$ = 0.0139(7) |
| | | | | | | $U_{eq}$ = 0.017 |
| Se(1) | 2*c* | 1.0 | 0 | 1/2 | 0.38833(18) | $U_{11}$ = 0.0208(10) |
| | | | | | | $U_{33}$ = 0.0187(19) |
| | | | | | | $U_{eq}$ = 0.020 |
| Se(2) | 2*c* | 1.0 | 0 | 1/2 | 0.81586(18) | $U_{iso}$ = 0.0172(6) |
| $R_{wp}$ | | | | 9.02% | | |
| $R_B$ | | | | 2.11% | | |
| GOF | | | | 5.85 | | |